Exploring the physics of the relativistic energy-momentum relationship

A.C.V. Ceapa

PO Box 1-1035, R-70700 Bucharest, Rumania

E-mail: alex_ceapa@yahoo.com

Considerations on the complementary time-dependent coordinate transformations emboding Lorentz transformation (LT) show that the relativistic energy-momentum relationship, implicitly the relativistic mass and energy, do not depend on the β appearing in LT, being associated to the absolute motion of a particle and related to its inner structure. Results concerning the concept of operational theory and its application to the electromagnetic and gravitational field theories, as well as to the quantum mechanics are given in appendixes.

It is interesting to examine in more detail the quantity

$$p^\mu p_\mu = E^2/c^2 - p^2, \quad (1)$$

where

$$E = mc^2, \quad p = mv \quad (2)$$

are, respectively, the relativistic energy and linear momentum of a free particle of relativistic mass

$$m = \beta m_o, \quad (3)$$

rest mass $m_o$ and velocity v, in relation to the laws of transformation of the energy

and linear momentum of a free particle under the coordinate transformation equations

x'=x-vt, y'=y, z'=z, t'=t-vx/c²     (4)

and

x'=β(x-vt), y'=y, z'=z, t'=β(t-vx/c²),     (5)

called complementary time dependent coordinate transformations.

## 1. Complementary Time Dependent Coordinate Transformations

We distinguish between ordinary time dependent coordinate transformations (OTs) and complementary time dependent coordinate transformations (CTs). The OTs are simply obtained by changing angles and lengths in time independent coordinate transformations into time dependent quantities. They are represented by spatial rotations and translations. CTs are related to the tracing of radii vectors by physical signals traveling through space with constant velocity υ. This tracing is required by our need of knowing the length and the direction of the radius vector of any geometrical point belonging to a "stationary" subspace before drawing and projecting it onto the coordinate axes of a stationary coordinate system (K) in space which is at absolute rest (see also Sect.1.1 in ref.1) as long as such points are aimed by an uniform translatory motion. The CTs are established for points of a subspace coinciding at an instant of time with space points. Unlike OTs which can be written whenever after the radii vectors of a geometrical point were traced by a pencil, the CT can be written only after the radii vectors we trace by a pencil have previously been traced by physical signals of identical nature. Depending on the nature of the physical signals tracing radii vectors, we have a CT or another. For light signals we have LT as a particular CT in the three-dimensional space. The preference for LT is related to the large value of c in comparison with the speeds of all the known physical signals, to the propagation of the electromagnetic and gravitational fields at speed c, but especially to the fact, pointed out elsewhere[2], that c is also a subquantum velocity. The equations of any CT are those of LT with c changed to the speed υ of the used physical signal. Specific to all CTs is their time equation obtained in their preliminary form as the time equivalent of their spacial equation written along the direction of motion of k relative to K. The manner in which we use the physical signals to establish a CT is just that used to obtain LT in Sect.4 of ref.1. Like LT, any CT reduces to GT in the "low-velocity" approximation. This only means that in such a situation OO' becomes negligible in comparison with $OP_1$ (OP') and $O'P_1$ (O'P') in the diagram in Fig.5 (10) in ref.1, ct* reduces to ct and, implicitly, t* reduces to the time t on the time axis. As concerns the homogeneity of the CTs, it originates in the initial superposition of the coordinate systems k and K required to obtain the geometry in Figs.5 and 10. The most simple CT is that given by Eqs.(38) in ref.1. It follows from the first of Eqs.(5) and (21) related to the upper diagrams in Figs.1 and 2 in ref.1. The raising of Eq.(21) was largely discussed in Sect.4.1. Like LT, Eqs.(38) form a group.

For v=c, Eqs.(38) reduce to

x'=x-ct, t'=t-x/c.    (6)

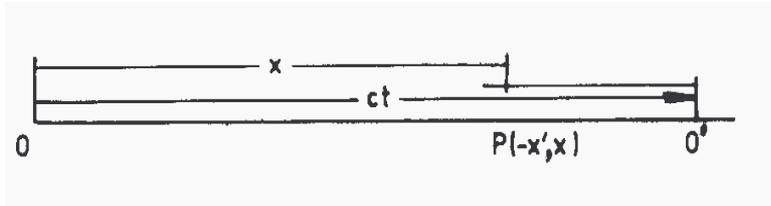

Fig.1

Eqs.(6) are related to the diagram in Fig1. Since k is carried by the tip of a light signal, only geometrical points P(x',x)∈ (O',O), where O' and O are, respectively, the origins of k and K, can be joined by light signals. Naturally, Eqs.(6) do not form a group; this because, carried by light signals leaving simultaneously O, the coordinate systems $k_A$ and $k_B$ are always superposed to each other. Moreover, the time component of Eqs.(6) should not be identified with the time relation

t'=t-r/c

which, connecting two synchronous clocks, does not belong to a coordinate transformation (for consequences of CT see Appendixes 1 to 3 below).

## 2. Transformation Laws for Energy and Linear Momentum

Assume for the beginning that we do not know that the energy and the linear momentum form a four-vector. Also assume that we do not know the transformation laws satisfied by the covariant and contravariant components of a four-vector. So that we propose to establish the transformation laws of the two from the invariance of the action

E't'-p'x'=Et-px    (7)

under Eqs.(4) and (5), connecting the coordinate system K at absolute rest to the parallel coordinate system k in uniform rectilinear motion along the common x', x axis of coordinates. Denote by E, p and E', p' the energies and linear moments of a free particle in relation to K and k, respectively. Substituting Eqs.(4), (5) and their inverses in Eq.(7), we get, respectively, the equations

E=E'+p'v, p=p'+E'v/$c^2$,    (8')

$E=\beta(E'+p'v)$, $p=\beta(p'+E'v/c^2)$     (8")

and

$E'=E-pv$, $p'=p-Ev/c^2$,     (9')

$E'=\beta(E-pv)$, $p'=\beta(p-Ev/c^2)$.     (9")

Eqs.(8) and (9) constitute the searched laws of transformation of the energy and linear momentum under the CT Eqs.(4) and (5). Each of these laws is analogous to the inverse of the CTs taken into account as a consequence of the last.

### 3. Contravariant and Covariant Four-Vectors

It is well-known that the contravariant and covariant components of a four-vector, respectively $A^\mu$ and $A_\mu$, are mathematically given by the transformation laws[3]

$A^\mu = (\partial x^\mu / \partial x'^\nu) A'^\nu$, $A_\mu = (\partial x'^\nu / \partial x^\mu) A'_\nu$,     (10)

where Greek indices run from 0 to 3, with the coordinates $x'^\mu$ and $x^\mu$ connected by LT. The derivation of the transformation laws of the contravariant and covariant components $p^\mu$ and $p_\mu$ of the four-momentum from the invariant called action in Sect.2 makes explicit the way in which the mixture of times and coordinates in the LT equations raises Eqs.(10). Continuing this line of thought, we further consider a physical quantity which is a differential function of $x'$, $x'_o(=ct')$ that in their turn, by the LT equations

$x'=\beta(x-vx_o/c)$, $x'_o=\beta(x_o-vx/c)$,

are continuous functions of $x$, $x_o(=ct)$ with partial derivatives.

The differential of this function is

$df=(\partial f/\partial x)dx+(\partial f/\partial x_o)dx_o=$

$=[(\partial f/\partial x')(\partial x'/\partial x)+(\partial f/\partial x'_o)(\partial x'_o/\partial x)]dx+$

$+[(\partial f/\partial x')(\partial x'/\partial x_o)+(\partial f/\partial x'_o)(\partial x'_o/\partial x_o)]dx_o=$

$=\beta[\partial f/\partial x'-v/c)(\partial f/\partial x'_o)]dx+\beta[-(v/c)(\partial f/\partial x')+\partial f/\partial x'_o]dx_o$.

With the notations

$\partial f/\partial x=A$, $\partial f/\partial x_o=A_o$, $\partial f/\partial x'=A'$, $\partial f/\partial x'_o=A'_o$,

we regain the first of Eqs.(10). This result is worthwhile because it infers that **the components of any four-vector are always derivatives of a function which**

**must be identified for its physical meaning and consequences to be well-determined**. Unfortunately, there is the common tendency of endowing the four-vectors with a mysterious physical existence which, by their transformation law analogous with LT, extends onto the last.

### 4. Four-Momentum, Proper Frame

The four-momentum was defined by[3]

$p^\mu = m_o c u^\mu$ ,

where $u^\mu = dx^\mu/ds$ is the four-velocity, $ds = (\eta_{\mu\nu} dx^\mu dx^\nu)^{1/2}$ is the metric of the Minkowskian space and $\eta_{\mu\nu} = (-1,-1,-1,+1)$ is the suitable metric tensor. When written with respect to a coordinate system K at absolute rest (see also Sect.3.2 in ref.1), for which $ds = \beta^{-1} c dt$, the four-momentum is given by

$p^\mu = m_o \beta\, dx^\mu/dt = (m_o \beta v, m_o \beta c)$,

in agreement with the classical definition of the linear momentum and the dependence on velocity of the mass. When written with respect to the "stationary" coordinate system k' in which a particle is at rest (v=0)-called proper frame, the four-momentum takes the preliminary form $p^\mu = m_o dx^\mu/d\tau$ by virtue of $ds = c d\tau$, where $\tau$ is the proper time, and a final form $p^\mu = (m_o \beta v, m_o \beta c)$, identical to that relative to K, by the equation $dx = v dt = v \beta d\tau$ , following from the standard LT equations under the condition dx'=0 required to measure $d\tau$ . Thus, against the appearances, we obtain the natural result that **whenever a free particle moves with respect to K with constant velocity v or is at rest with respect to a coordinate system moving with the same velocity relative to K (its proper frame), it possesses the same mass $m_o\beta$, the same energy $m_o\beta c^2$ and** (although we cannot define a non-zero velocity in this case) **the same quantity of motion. Stating that the mass and the energy of a particle are, respectively, $m_o$ and $m_o c^2$ in its proper frame is false and misleading as long as that particle is carried by its proper frame**. The values $m_o$ and $m_o c^2$ are true only for a particle at rest in a stationary coordinate system. If Einstein connected these values to the proper frame, he did it only because missing the meaning of $\Xi$ in his original paper on relativity (see also Sect.3.7 in ref.1), and believing that he eliminated the coordinate system at absolute rest from his theory of "relativity", he was compelled to introduce the concept of proper frame just as he was compelled to extend the L-principle to "stationary" coordinate systems. Thus, whenever we use the proper frame we must keep in mind that the true quantities defining a particle at rest with respect to it are a non-zero quantity of motion, a mass $m = m_o \beta$ and an energy $E' = m_o \beta c^2$ (here $\beta$ having nothing in common, as concerns its origin, with $\beta$ occurring in the Lorentz transformation!). In fact, the quantities $m_o\beta$ and $m_o\beta c^2$ are always associated to the absolute motion of a particle. This can be explained by that any state of motion of a particle alters its subquantum basic state.

### 5. The Relativistic Energy-Momentum Relationship

Let us write Eqs.(8) and (9) in relation to the proper frame of a free particle. Assuming p'=0, they are

$E=E'$,   (11')

$E=\beta E'$,   (11'')

and

$E'=\beta^{-2}E$,   (12')

$E'=\beta^{-1}E$.   (12'')

The last of Eqs.(8) and (9) are

$|p|=Ev/c^2$.   (13)

Since Eqs.(4) and (5) [the inverses of Eqs.(4) and (5)] connect a coordinate system k (K) in uniform rectilinear motion with respect to a coordinate system K (k) at absolute rest, whenever k represents a moving (rest) proper frame, the energy E' appearing in Eqs.(11)[(12)] (see also Sect.4) is

$E'=\beta m_o c^2$ $[E'=m_o c^2]$.

Thus Eqs.(11) and (12) become

$E=\beta m_o c^2$,   (14')

$E=\beta^2 m_o c^2$   (14'')

and

$E=\beta^2 m_o c^2$,   (15')

$E=\beta m_o c^2$.   (15'')

The quantity (1) reduces by Eq.(13) to $E^2/c^2-p^2=\beta^{-2}E^2/c^2$. Further, by Eqs.(14) and (15) it takes the forms

$E^2/c^2-p^2=m_o^2 c^2$   (16')

$E^2/c^2-p^2=m^2 c^2$.   (16'')

We recognize in Eq.(16') the standard relativistic energy-momentum relationship. We also see that Eq.(16"), which is $\beta^2$ times Eq.(16') and embodies a change of origin on the energy scale, has previously been missed by assuming that $E'=m_oc^2$ for particles at rest in their proper frames, irrespective of the state of rest or uniform translatory motion of the last.

Therefore, obtained by Eqs.(14) and (15) as well, Eqs.(16) do not depend on the presence of β in the CT taken into account. Implicitly, **the dependence on β of Eqs.(2) and (3) is**, **in accord with the experiment**, **not determined by LT**. The coincidence of Eq.(16') [(16")] with that resulting from Eqs.(2) and (3) for a free particle moving relative to a K at absolute rest [in uniform rectilinear motion] assures the invariance of $p^\mu p_\mu$ in relation to LT.

## 6. The True Derivation of Standard Energy-Momentum Relationship

The true derivation of the standard energy-momentum relationship is related to a particle at absolute rest with respect to a stationary coordinate system K. Its suitable energy is $E=m_oc^2$. Its linear momentum is p=0. Inserting these values in Eq.(8") we obtain

$p'=-E'v/c^2$, $E=\beta^{-1}E'$.

Thus the energy and the linear momentum of this particle relative to a coordinate system k in uniform translatory motion with respect to K, as well as those of a particle moving with the same velocity relative to K, are $E'=\beta m_oc^2$, $|p'|=\beta m_ov$. The relationship (16') is immediate.

Observe that, by replacing the stationary coordinate system K by a "stationary" coordinate system K, and denoting the energy and the linear momentum of a free particle respectively by $mc^2$ [with m given by Eq.(3)] and p=0, we also deduce by Eq.(8") the energy-momentum relationship (16"). Unlike their derivation by means of Eqs.(11) and (12), Eqs.(16) have now a precise physical significance.

## Appendix 1: Four-Dimensional Consequences of CTs

The LT equations, as well as the equations of any other CT defined in Sect.1, predict a four-dimensional metric and, implicitly, a four-dimensional space-time physically determined but with no physical significance. When the physical signals are light signals, the space-time is just that of Minkowski. There results that **the Minkowski space-time is a consequence of the tracing by light signals of radii vectors of geometrical points belonging to moving subspaces**. By the four-quantities and invariants also having their origin in the tracing of radii vectors by light signals, **the Minkowski space-time appears to be a rigorous framework to describe the physical reality filling space**. The events are determined in relation to four-dimensional coordinate systems. Beside the Minkowski space-time there is the four-space, also formal, associated to the four-momentum $p^\mu=(p,E/c)$ just as the former was attached to the four-vector $x^\mu=(x,ct)$. This is the energy-momentum space which Caianiello joined[4] with the Minkowski space-time into an eight-dimensional space, and which metric enabled him to deduce the maximum acceleration $a_M$. The endowing of phase space with a metric just expresses the physical determination of the quantum theory. Other aspects to be noted concern the metric of the Minkowski space-time itself. Thus, since t* denotes the time in which a physical signal traces the radius vector of a moving geometrical point P, ct does not define the x of P as a function of time but just as a length. Therefore, we can not write dx=cdt, but either dx=vdt or $(dx^2+dy^2+dz^2)^{1/2}=vdt$. It is for this reason that the metric s, reduced to $(y^2+z^2)^{1/2}$ or 0 by x=ct, was used by Einstein only in connection with the light cone, while the infinitesimal metric ds (which always differs from zero) to get the main predictions of his theory.

### Appendix 2: Operational Theories

Whenever physicists searched for solutions of an equation or a set of equations mapping to a less understood physical reality, they resorted to additional mathematical constraints on the quantities related by those equations to reach their goal. Often the solutions of these equations were in accord with experiments. Sometimes they were not, and the confidence in them diminished and delayed clear experimental results. The last is the case with the weak gravitational waves predicted by Einstein's theory of general relativity on the analogy with the plane electromagnetic waves. Like transverse waves, they were mathematically defined by imposing the transverse-traceless conditions[5]

$$\Psi^\mu_{\ \nu} u^\nu = 0, \quad \Psi^\mu_{\ \mu} = 0 \quad (17)$$

to their potentials

$$\Psi^{\mu\nu} = h^{\mu\nu} - (1/2)\eta^{\mu\nu} h^\sigma_{\ \sigma},$$

where $h^{\mu\nu}$ are the infinitely small components of the metric tensor

$$g^{\mu\nu} = \eta^{\mu\nu} + h^{\mu\nu}$$

and $u^\nu$ is the four-velocity of the wave, that satisfy Einstein's linearized equations of the gravitational field.

Nevertheless, we see that actually the transformation laws of $\Psi^{\mu\nu}$ are implied by

the transformation laws

$$h'^{\mu\nu}=(\partial x'^{\mu}/\partial x^{\alpha})(\partial x'^{\nu}/\partial x^{\beta})h^{\alpha\beta} \qquad (18)$$

of $h^{\mu\nu}$ under the inverse of the CT Eqs.(6) that connect a coordinate system moving with velocity c with respect to one at absolute rest. By a little algebra, Eqs.(18), (6) and (17) show that the only non-zero potentials of a weak gravitational wave traveling along the x axis are

$$\Psi'^{2}{}_{2}\equiv\Psi^{2}{}_{2}=-\Psi^{3}{}_{3}\equiv-\Psi'^{3}{}_{3}, \quad \Psi'^{2}{}_{3}\equiv\Psi^{2}{}_{3}=-\Psi^{3}{}_{2}\equiv\Psi'^{3}{}_{2}, \quad (19)$$

i.e., just those deduced by the traceless condition mathematically imposed. That is to say, Eqs.(18) connect wave potentials defined in relation to both the coordinate system carried by the wave and the stationary coordinate system of an observer recording the wave.

Going on, since the quantities characterizing electromagnetic and gravitational fields are attached to geometrical points of subspaces traveling at speed c through empty space, and the CT Eqs.(6) refer points of such subspaces to coordinate systems at rest in this space, we focus our attention on the theories of electromagnetic and gravitational fields which exhibit retarded quantities for revealing the mathematical constraints historically imposed to obtain them[6].

Consider for the beginning the mathematical quantities f and $\xi^{\mu}$ appearing, respectively, in the electromagnetic and gravitational theories[7] by the gauge transformation of their four-potentials. Observe that the dependence on t-x/c of f and $\xi^{\mu}$ has been obtained by imposing the Lorentz condition

$$\partial A^{\mu}/\partial x^{\mu}=0$$

and its gravitational counterpart

$$\partial \Psi^{\mu\nu}/\partial x^{\nu}=0$$

on the four-potentials of the plane electromagnetic and gravitational waves, respectively, $A^{\mu}$ and $\Psi^{\mu\nu}$. Since $A^{\mu}$ and $\Psi^{\mu\nu}$ are defined in a coordinate system k moving with velocity c with respect to a coordinate system at absolute rest-the working coordinate system of the observer-their time dependence, as well as that of f and $\xi^{\mu}$, follows from the second of Eqs.(6) just as a result of the velocity of k relative to K.

Moreover, because f and $\xi^{\mu}$ belong to the mathematical basis of the two theories, the time dependence previously imposed on them is equivalent to the more general requirement, namely that these theories are developed in the working coordinate systems of the observers performing measurements of physical quantities.

Therefore, the last of Eqs.(6) accounts for the retarded potentials, whose omnipresence has been until now only in agreement with experiment[8], in that they are defined in coordinate systems k traveling at speed c with respect to a K in empty space and measured by observers with respect to such systems by the diagram in Fig.4 in ref.1.

Extending these conclusions about the role played by Eqs.(6) in founding the two theories to the full class of complementary time dependent coordinate transformations (i.e., to those obtained by other physical signals than light), we get the concept of operational theory stated as[6,9]:

**A physical theory is an operational theory if and only if the quantities entering into its equations are expressed in the working coordinate systems of the observers performing measurements.**

The main consequence of this concept is that the modern physics becomes a system of operational theories valid in the working systems of the observers performing measurements. This means implementing the new theories as operational ones and removing from the older theories those mathematical constraints historically imposed by the process of knowledge exclusively for obtaining the time dependence of the physical quantities required by experiment.

Both the definition of $p^\mu$ investigated in Sect.4, and the subsequent rewriting of the equations of the relativistic quantum mechanics in a moving $K_i$ (pointed out in the next Appendix), as well as the new derivation and meaning of the Hubble law in relation to Eqs.(4)[6,10], which applied to the furthest quasars recently discovered does not support the receding of the galaxies from the Earth, represent first steps done in this aim. In addition, by obtaining the potentials (19) of a weak (plane) gravitational wave[6,11], and converting the equation

$$t'=t(1+2\phi/c^2)^{1/2}(1-v^2/c^2)^{1/2},$$

which predicts the effect that a Newtonian gravitational field of potential $\phi$ has on an elapsing of time, to[12]

$$t'_K=t(1+2\phi/c^2)^{1/2}\cong t(1+\phi/c^2),$$

by the last of Eqs.(6) (t' being defined in the proper frame of the field and $t'_K$ in the coordinate system of a moving observer), and not by imposing prior mathematical constraints (e.g., $dx^i=0$ in the metric[12]), the operational approach also extends to the weak field approximation of general relativity.

### Appendix 3: Absolute Coordinate Systems

A main result obtained in Sect.4 was that the energy and the quantity of motion of a particle at rest with respect to its proper frame moving with constant velocity v relative to K are given by Eqs.(2) without an observer isolated in that proper frame to be able to identify them by calculation or measurements. We concluded that (see also ref.6) the energy $m_o c^2$ is specific to a particle at rest with respect to a coordinate system at absolute rest. It appears that by their defining relationships the "relativistic" energy and momentum were from the beginning estimated in relation to a coordinate system at absolute rest (we here denote it by $K_{abs}$).

Since the invariant $p^\mu p_\mu$ predicts the standard energy-momentum relationship (6'), the equations of the relativistic quantum mechanics and, implicitly, the coupling constant $m_o c^2$ between the systems of subquantum particles they involve, were defined in relation to $K_{abs}$. But as the invariant $p^\mu p_\mu$ was estimated in relation to the proper frame of the particle, and the energy and momentum of the last are given by Eqs.(2) and (3), the equations of the relativistic quantum mechanics, involved by Eq.(16") with $m_o$ changed to $m_i=\beta m_o$ and the suitable coupling constant $m_i c^2$ are defined in relation to the coordinate systems $K_i$ moving with velocities $v_i$ relative to $K_{abs}$, as well. And, since the observers are associated to $K_i$ not to $K_{abs}$, the writing of these equations in $K_i$ corresponds to the foundation of the relativistic quantum mechanics as an operational theory, in full agreement with the definition of such a theory given in Appendix 2.

Concerning the new coupling constant $m_i c^2$ between the systems of subquantum particles[2], it reveals an extra-coupling between these systems and the state of motion of a quantum particle defined by $v_i$. It is this extra-factor that which predicted by the energy-momentum relationship is related to the existence of an absolute coordinate system in physics despite our impossibility of identifying it in nature as a system of reference bodies.

The usual work with Eq.(16') and, implicitly, with the equations of the relativistic quantum mechanics defined in relation to $K_{abs}$ mathematically originates in our possibility to drop $\beta^2$ in Eq.(16"), while physically in that the coordinate systems $K_i$ of the observers can actually be related to the stationary coordinate systems $\Xi_i$ as shown in Sect.3.6 in ref.1.

### References

[1] A.C.V. Ceapa, General Physics/9911067.

[2] A.C.V. Ceapa, Physical Grounds of Einstein's Theory of Relativity. Roots of the Falsification of 20th Century Physics. (3rd Edition, Bucharest, 1998), Part III.

[3] L.D. Landau and E.M. Lifshitz, The Classical Theory of Fields (Pergamon, N.Y.,